\documentclass[allclo]{FBSart}
\runningauthor{V.D. Efros}
\runningtitle{Elimination of Rotational Degrees of Freedom}
\newcommand{\re}{\ref}
\newcommand{\be}{\begin{equation}}
\newcommand{\ee}{\end{equation}}

\newcommand{\la}{\label}
\newcommand{\ber}{\begin{eqnarray}}
\newcommand{\eer}{\end{eqnarray}}

\begin{document}

\title{Elimination of Rotational Degrees of Freedom
 in Expansion Methods for Three Nucleons}

\author{V.D. Efros}
\institute
{
Russian Research Centre "Kurchatov Institute",
Kurchatov Square 1,\\ 123182 Moscow, Russia\thanks{\textit{email:}
efros@mbslab.kiae.ru}}


\begin{abstract}
Euler angles determining rotations of a system as a whole
are conveniently separated in three--particle basis functions.
Analytic integration 
of matrix elements over Euler angles is done in a general form.
Results for the 
Euler angle integrated
matrix elements of a realistic NN interaction are listed.
The partial wave decomposition of correlated three--body states
is considered.
\end{abstract}

\maketitle

\bigskip

\section{Introduction}
When expansion methods are applied 
in three--particle problems
matrix 
elements (ME) are to be calculated which are six--dimensional integrals.
It is expedient to choose Euler angles, representing
rotations of a system as a whole, as three of the six
degrees of freedom and to carry out the integration over Euler angles
analytically. Then one is left with integrations that are three--dimensional 
only. Such low--dimension integrations can very efficiently be done with 
regular quadratures.
The elimination of rotational  degrees of freedom has been used by
several groups \cite{del,ki,elot00} at solving both the three--nucleon
Schr\"odinger dynamic equation and that of the method of integral transforms
\cite{efr85}. No restrictions on the form of basis functions arise in such an
approach, and in conjunction with use of an appropriate set of basis
functions the techniques are probably the fastest and easiest ones to solve
most of the nuclear three--body problems of  interest. Therefore, 
it seems useful to list the formalism. In Sec. 2 the 
separation of Euler angles  
is  performed for 
general form three--particle functions. In Sec. 3 analytic integration 
over Euler angles
of general type ME is carried out which is applied 
in Sec. 4 to 
calculate the ME pertaining to a realistic NN interaction. The partial wave
decomposition of correlated basis functions is  considered in Sec. 3 as well.
Sec. 5 contains
comments on symmetry properties of Euler angle integrated ME.

In Ref. \cite{dav} most of the ME considered in Sec. 4 were also
calculated, however, our techniques are different. 
Both the results and techniques of the present paper
are substantially simpler. 
Besides, we present the general formula for Euler angle integrated ME, and
our choice of the body reference frame 
leads to simplifications. 
 
Contrary to the correlated basis functions case, ME of
one--body and two--body operators can be calculated in a  simple way
without the Euler angle integration when e.g. uncorrelated 
hyperspherical harmonics are used as a basis set. However, the
present techniques are helpful also in the latter case when ME
of NNN force or those involving two--body subsystems 
at solving reaction problems  are considered. 

While the formulae are written down below for the case of the 
three--nucleon problem 
they may be applicable also beyond these
frames.
 
\section{Separation of Euler Angles}

We use the Jacobi vectors
\be
{\vec u}=\frac{1}{\sqrt{2}}\left({\vec r}_2-{\vec r}_1\right),\qquad{\vec v}=
\sqrt{\frac{2}{3}}\left({\vec r}_3-\frac{{\vec r}_1+{\vec r}_2}{2}\right)
\ee
where ${\vec r}_i$ are nucleon positions,
and we consider three--dimensional rotations
${\vec u,\vec v}\rightarrow{\vec u',\vec v'}$.
We interpret rotations as rotations of a coordinate system so that
a rotation ${\vec a}\rightarrow{\vec a'}$ means that
$\{a_x,a_y,a_z\}\equiv{\vec a}$ and $\{a_x',a_y',a_z'\}\equiv{\vec a'}$
are coordinates of a vector
in the old
and new coordinate systems, respectively.
Spatial components $F_{LM}$,   with
given total orbital momentum and its projection quantum numbers,
of three--particle basis functions
transform under rotations as \cite{var}
\[
F_{LM'}({\vec u',\vec v'})=
\sum_{M=-L}^L D^L_{MM'}(\omega)F_{LM}({\vec u,\vec v})
\]
where $\omega$ denotes collectively three Euler angles 
parameterizing a rotation
and $D^L_{MM'}$ are the Wigner D--functions.

In order to separate the Euler angle dependence of 
$F_{LM}({\vec u, \vec v})$ we
consider rotations ${\vec u}'={\vec u}'({\vec u},{\vec v})$, 
${\vec v}'={\vec v}'({\vec u},{\vec v})$ which are different for different 
${\vec u}$, ${\vec v}$ vectors,  so that  $\omega=\omega({\vec u,\vec v})$. 
We
use the inverse
relation
\be
F_{LM}({\vec u, \vec v})=
\sum_{M'=-L}^L D^{L*}_{MM'}(\omega)F_{LM'}({\vec u',\vec v'}).
\la{rot}
\ee
At given ${\vec u}$ and ${\vec v}$, we choose
the new coordinate system in a way that it corresponds to a body
reference frame
associated with the
${\vec u,\vec v}$ plane. 
The corresponding body--frame coordinates $u'_i,v'_i$ may be expressed
in terms of scalars $u$, $v$, and $({\vec u}\cdot{\vec v})$. 
Treated in this way, Eq. (\ref{rot}) becomes
an expansion of $F_{LM}({\vec u,\vec v})$ over D--functions depending on 
Euler angles  
with coefficients depending
on scalar variables. The Euler angles  determine positions
of the body reference frame with respect to a laboratory reference frame.

In particular, a convenient
choice \cite{efr73} is to
direct the $z$--axis of the body reference frame along the vector
${\vec u}$, and to place the $x$--axis of the body reference frame in
the ${\vec u,\vec v}$ plane in a way that the projection of the vector
${\vec v}$
onto this axis is positive. Then one has
\be
u_x'=0,\quad u_y'=0,\quad u_z'=u,\quad v_x'=v\sqrt{1-t^2},\quad v_y'=0,
\quad v_z'=vt\la{poi}\ee
where $t=(\bar{\vec u}\cdot\bar{\vec v})$, $\bar{\vec u}$ and 
$\bar{\vec v}$ being
the unit vectors in the directions of ${\vec u}$ and ${\vec v}$.
As a result, one comes to the expansion
\be
F_{LM}({\vec u,\vec v})=\sum_{M'=-L}^L D^{L*}_{MM'}(\omega)
\underline{F_{LM'}}(u,v,t)
\la{rot1}
\ee
where  the Euler angles $\omega$ parametrize the above defined
rotation into the body reference frame, and the notation 
$\underline{F_{LM}}(u,v,t)$ stands for the quantities\\ 
$F_{LM}({\vec u',\vec v'})\equiv
F_{LM}(u_x',u_y',u_z',v_x',v_y',v_z')$ taken at the space point 
from (\re{poi}):
\be
\underline{F_{LM}}(u,v,t)\equiv
F_{LM}(u_x'=0,u_y'=0,u_z'=u,v_x'=v\sqrt{1-t^2},v_y'=0,v_z'=vt).\la{bar}
\ee
Writing down the formulae we shall consider $\underline{F_{LM}}$ to 
be real which is commented below in connection with Eq. (\re{emod}).

The coordinates $\omega\equiv\{\alpha,\beta,\gamma\},u,v,t$ 
ranging between their natural boundaries
realize a mapping onto 
the whole phase space
$({\vec u},{\vec v})$. The correspondence between the points in the phase space
and the $\alpha,\beta,\gamma,u,v,t$ coordinates is a one--to--one correspondence everywhere
except for the exceptional sub--areas where  
the ${\vec u,\vec v}$ plane is undefined and hence
the body reference frame cannot be  
defined. These are sub--areas where
vectors ${\vec u}$ and ${\vec v}$
are collinear, or $u=0$, or $v=0$. Many sets of Euler angles correspond
to the same point in the phase space if the point belongs to these sub--areas.  
For the case of integrands which are non--singular
in the exceptional sub--areas the contribution of these
sub--areas to ME vanishes (not speaking of the fact that the phase space
volume element suppresses the contributions of $u=0$ and $v=0$ sub--areas,
see Eq. (\re{tau}) below). Therefore, one can transform the integration 
over ${\vec u},{\vec v}$ into the integration over $\alpha,\beta,\gamma,u,v,t$ 
and apply Eq. (\ref{rot1}).\footnote{In addition, it is easy to see that 
Eq. (\ref{rot1}) remains valid in the exceptional sub--areas where 
${\vec u}$ and ${\vec v}$ are collinear, and where $v=0$. Both sides
of  Eq. (\ref{rot1}) turn to $Y_{LM}(\beta,\alpha)f(u,v)$ in these sub--areas,
and integration over them using the coordinates $\alpha,\beta,\gamma,u,v,t$
leads to correct results.} 

Our basis functions $F_{LM}$ are regular single--valued functions of
the points of the phase space. This is in contrast to D--functions themselves
which
do not possess such property in the above mentioned exceptional sub--areas.
While this does not cause problems when Eq. (\ref{rot1}) is used for 
 integration purposes, use of products of D--functions and functions of 
internal variables as basis functions would be a 
disadvantage.\footnote{At given
$L$ and $M$ values D--functions
form a complete set of $2L+1$ functions of three coordinates determining
the position of a system as a whole.  Constructing the basis set $F_{LM}$
one, for example, can use   another set of $2L+1$ 
functions instead of D--functions
which have the same property but which are basically polynomials
of the components of the vectors ${\vec u},{\vec v}$ and are 
regular in the whole
phase space. These functions are to be multiplied by functions of three
internal variables. 
This basis proved to be very efficient in atomic physics 
\cite{efrzh}. 

When such a basis is used to calculate ME of
some simple operators there is no need to apply Eq. (\ref{rot1}).
The integrands in the left--hand side
of Eq. (\re{ov}) below can be decomposed in a closed form
into the sum of components with given
$L$ values. 
Only the $L=0$ component contributes to the right--hand side of (\re{ov}),
and no summation over $M''$ is then required, see \cite{efrzh}.}    

Let us also comment on 
the aspect of symmetry with respect to particle permutations.
A well--known way to get totally antisymmetric three--nucleon
basis functions is to combine spatial basis functions
that are components of irreducible
representations of the three--particle permutation group with 
spin--isospin functions of conjugated symmetry.   
The corresponding spatial basis functions
are often constructed via application of 
symmetrization
operators to complete sets of initial "simple" functions.
It is convenient,
and sufficient,
to use the expansion (\ref{rot1}) of symmetrized functions
involving one and the same body reference 
frame described above. In contrast to the basis functions themselves, 
this reference frame does not possess symmetry properties with respect to 
permutations of three particles. 
The arising symmetrized functions $F_{sym}$ are 
expressed in terms of initial functions $F$ with 
the help of relations of the type
\be F_{sym}({\vec u},{\vec v})=\sum_PC_PF(P{\vec u},P{\vec v})\la{symm}\ee
where $P$ are particle permutations, and $C_P$ are given coefficients. Hence,    
the  required body--system components (\re{bar})
of symmetrized functions are obtained with the help of such type relations 
applied at the  points $({\vec u}',{\vec v}')$
given by Eq. (\re{poi}). 
One thus needs to form linear combinations of the corresponding vectors
\[P{\vec u}'=a_P{\vec u}'+b_P{\vec v}',
\qquad P{\vec v}'=c_P{\vec u}'+d_P{\vec v}'\] 
and to calculate initial functions at the  points $(P{\vec u}',P{\vec v}')$.
When, alternatively, the basis functions antisymmetric with respect to nucleon 
permutations
are obtained via a direct 
antisymmetrization of angular momentum coupled
products of space functions and spin--isospin
functions the procedure is similar.

ME between functions of the form (\re{rot1})
will be calculated below. The volume element is
\be
 d\tau=d{\vec u}d{\vec v}=d\omega d\tau_{int},\qquad\
 d\tau_{int}=u^2v^2dudvdt.
\la{tau}
\ee
Here the 
relationship $d\bar{{\vec u}}d\bar{{\vec v}}=d\omega dt$, see Appendix,
is used. 
The overlap--type integrals are
\ber
\int d\tau G_{L'M'}^*\hat{O}_{u,v,t}F_{LM}=
\delta_{LL'}\delta_{MM'}8\pi^2(2L+1)^{-1}\nonumber\\
\times\int d\tau_{int}\sum_{M''=-L}^L\underline{G_{LM''}}(u,v,t)
\hat{O}_{u,v,t}\underline{F_{LM''}}(u,v,t).\la{ov}
\eer
Here $\hat{O}_{u,v,t}$ is an operator acting on $u,v,t$, and
the orthogonality properties (\ref{aor}) are used.

\section{Matrix Elements of Tensor Operators}

To calculate ME it is expedient to decompose the operators 
into sums of tensors irreducible with respect to rotations. 
Performing integration over Euler angles
analytically is then 
possible in practically important cases. The following relationships
are useful to this aim.

1. Let $\hat{O}_{k\kappa}$ be the irreducible tensor operator \cite{var} 
of the rank $k$.
We obtain its Euler angle integrated 
reduced ME $(G_{L'}||\hat{O}_{k}||F_L)$ defined as usual,
\[
\langle G_{L'M'}|\hat{O}_{k\kappa}|F_{LM}\rangle=
C_{LMk\kappa}^{L'M'}(2L'+1)^{-1/2}(G_{L'}||\hat{O}_{k}||F_{L}),\]
where $C_{LMk\kappa}^{L'M'}$ are the Clebsh--Gordan coefficients.
The result is
\be
\left(G_{L'}||\hat{O}_k||F_{L}\right)=8\pi^2(2L'+1)^{-1/2}
\int d\tau_{int}\sum_{M M' \kappa\atop M'=M+\kappa}
C_{LMk\kappa}^{L'M'}\,\underline{G_{L'M'}}\,
\underline{\hat{O}_{k\kappa}F_{LM}}.\la{gen}
\ee
Here $\underline{\hat{O}_{k\kappa}F_{LM}}$ means the quantity
$\hat{O}_{k\kappa}F_{LM}$ taken at the space point from (\re{poi}).
The relation (\re{gen}) is useful when 
one is able to calculate $\underline{\hat{O}_{k\kappa}F_{LM}}$
in a simple way.
The cases when the operator $\hat{O}_{k\kappa}$ includes multiplications by 
functions and differentiations fall into this category, cf. the next section. 

To get Eq. (\re{gen}) we define
the states
\be
 (OF)_{L_0M_0}^{kL}\equiv\sum_{\kappa+M=M_0}
C_{LMk\kappa}^{L_0M_0}\hat{O}_{k\kappa}F_{LM}. \la{co}
\ee
Since $\hat{O}_{k\kappa}$ is an irreducible tensor these states 
possess definite rotational quantum numbers $L_0$, $M_0$.  One  
therefore can expand them, 
as in Eq. (\re{rot1}), in terms of
products of D--functions and the
body--system functions 
$\underline{(OF)_{L_0M_0}^{kL}}(u,v,t)$ defined similar to
Eq. (\re{bar}).
Writing down the relation inverse to (\re{co}) and utilizing this
expansion gives
\[\hat{O}_{k\kappa}F_{LM}=\sum_{L_0M_0M'}C_{LMk\kappa}^{L_0M_0}
D^{L_0*}_{M_0M'}(\omega)\underline{(OF)_{L_0M'}^{kL}}.\]
Use of this relation, the D--function expansion of $G_{L'M'}$, and 
Eq. (\re{ov}) gives 
\[
\int d\tau G_{L'M'}^*\hat{O}_{k\kappa} F_{LM}=8\pi^2(2L'+1)^{-1}
C_{LMk\kappa}^{L'M'}\int d\tau_{int}\sum_{M''=-L'}^{L'}
\underline{G_{L'M''}}\,\,\underline{(OF)_{L'M''}^{kL}}.
\]
Substitution of the body--system version of Eq. (\re{co}) here,
\be
\underline{(OF)_{L'M''}^{kL}}=\sum_{M+\kappa=M''}
C_{LMk\kappa}^{L'M''}\underline{\hat{O}_{k\kappa}F_{LM}},\la{co1}
\ee
 gives Eq. (\re{gen}). 

More complicated ME can also be integrated over Euler angles. 
Let us comment   on e.g. the ME of the square of a 
three-nucleon Hamiltonian
$\hat{H}^2=(\hat{T}+\hat{V})^2$ entering the 
well--known Temple
lower bound for  binding energy.  
ME of $\hat{T}^2$ are calculated like
the overlap integrals (\re{ov}) with 
$\underline{F_{LM}}\rightarrow\underline{\hat{T}F_{LM}}$,
$\underline{G_{LM}}\rightarrow\underline{\hat{T}G_{LM}}$, ME of 
$\hat{T}\hat{V}+\hat{V}\hat{T}$ are calculated using the same replacements
and Eq. (\re{gen}), and ME of $\hat{V}^2$ are given by the sum of  
contributions of the form
\be\left\langle \left(G_{L_2}\otimes\chi_{S_2}\right)_{JM}|
\left(\hat{O}_{k_2}\cdot\hat\Sigma_{k_2}\right)\left(\hat{O}_{k_1}
\cdot\hat\Sigma_{k_1}\right)
|\left(F_{L_1}\otimes\varphi_{S_1}\right)_{JM}\right\rangle\la{v2}\ee
where $\hat{O}_{k}$ and $\hat\Sigma_{k}$ are
the rank $k=0$, 1, and 2 tensor operators,  
$(\hat{O}_{k}\cdot\hat\Sigma_{k})$ 
is their scalar
product \cite{var}, and  $\left(F_{L}\otimes\varphi_{S}\right)_{JM}$
symbolizes the $LSJ$ coupling.
The operators $\hat{O}_{k}$ and $\hat\Sigma_{k}$
act in the coordinate subspace and
the spin--isospin subspace, respectively,
and they are
irreducible with respect to rotations.
 
To calculate ME (\re{v2}) one performs recouplings, see Appendix,
leading to spatial and spin states of the type $(OF)_{LM}^{L_1k_1}$,
$(OG)_{LM}^{L_2k_2}$, $(\Sigma\varphi)_{SM_s}^{S_1k_1}$, and
$(\Sigma\chi)_{SM_s}^{S_2k_2}$ where the notation is similar to 
that of Eq. (\re{co}).
Subsequent application  of Eqs. (\re{rot1}) and (\re{ov}) 
to the states $(OF)_{LM}^{L_1k_1}$,
$(OG)_{LM}^{L_2k_2}$ gives the 
following expression for the
ME (\re{v2}), 
\be
\sum_{M_1\kappa_1M_2\kappa_2\atop M_1+\kappa_1=M_2+\kappa_2}
I(M_1,\kappa_1,M_2,\kappa_2)
J(M_1,\kappa_1,M_2,\kappa_2),\la{sv2}\ee
\[I(M_1,\kappa_1,M_2,\kappa_2)=\int d\tau_{int}\,
\underline{\hat{O}_{k_2\kappa_2}G_{L_2M_2}}\,\,
\underline{\hat{O}_{k_1\kappa_1}F_{L_1M_1}},\]
\begin{eqnarray*}J(M_1,\kappa_1,M_2,\kappa_2)=
8\pi^2(-1)^{S_1-S_2}\sum_{LS}(2S+1)
\left\{\begin{array}{ccc}
S & L & J\\L_1 & S_1& k_1\end{array}\right\}
\left\{\begin{array}{ccc}
S & L & J\\L_2 & S_2& k_2\end{array}\right\}\\
\times C_{L_1M_1k_1\kappa_1}^{L,M_1+\kappa_1}
C_{L_2M_2k_2\kappa_2}^{L,M_2+\kappa_2}
\left\langle (\Sigma\chi)_{SM_s}^{S_2k_2}|(\Sigma\varphi)_{SM_s}^{S_1k_1}\right\rangle.
\end{eqnarray*}
Here $M_s$ is arbitrary.

2. The
partial wave decomposition 
\ber
F_{LM}({\vec u,\vec v})=
\sum_{l_1l_2}Y^{l_1l_2}_{LM}(\bar{{\vec u}},\bar{{\vec v}})f_{l_1l_2L}(u,v),
\la{ef}\\
Y^{l_1l_2}_{LM}(\bar{{\vec u}},\bar{{\vec v}})=\sum_{m_1+m_2=M}
C_{l_1m_1l_2m_2}^{LM}
Y_{l_1m_1}(\bar{{\vec u}})Y_{l_2m_2}(\bar{{\vec v}})\la{y}
\eer
of correlated three--body states
could also help to calculate ME between them.
This applies e.g. to 
ME of a non--local NN interaction, or NN interaction with $l$-dependent
form factors.
To perform the decomposition we need to calculate the overlap integrals
\be
f_{l_1l_2L}(u,v)=\int d\bar{{\vec u}}d\bar{{\vec v}}
Y^{l_1l_2*}_{LM}(\bar{{\vec u}},
\bar{{\vec v}})F_{LM}({\vec u},{\vec v}).\la{po1}
\ee
If used directly, Eq. (\re{po1}) implies the
four--dimensional integration to
get $f$ at each $u,v$ point. But separating 
Euler angles and performing integration over them analytically 
one can replace
the
four--dimensional integration  with a single--dimensional
one. Indeed, similar to Eq. (\re{ov}),
using the Eq.~(\re{rot1})--type expansion for both factors in the
right--hand side
of Eq. (\re{po1}) we obtain  that
\be
f_{l_1l_2L}(u,v)=8\pi^2(2L+1)^{-1}\int_{-1}^1 dt
\sum_{M=-\lambda}^\lambda \underline{Y^{l_1l_2}_{LM}}(t)
\underline{F_{LM}}(u,v,t),\la{exp}
\ee
where $\underline{Y^{l_1l_2}_{LM}}(t)$ are the body--system values of
$Y^{l_1l_2*}_{LM}(\bar{{\vec u}},
\bar{{\vec v}})$,
\be
\underline{Y^{l_1l_2}_{LM}}(t)=A^{l_1l_2}_{LM}
Y_{l_2M}(\cos\theta=t,\varphi=0),\qquad
A^{l_1l_2}_{LM}=[(2l_1+1)/(4\pi)]^{1/2} 
C_{l_10l_2M}^{LM},\la{yb}\ee
and
$\lambda={\mathrm{min}}(l_2,L)$. 
 
If in the expansion (\re{ef}) the functions $f_{l_1l_2L}(u,v)$ are 
real\footnote{In the usual case 
of T--invariant Hamiltonians the
expansion coefficients are real for such a choice of basis functions.}
then
the body-system functions $\underline{F_{LM}}$ are real as well.
This is seen from the body--system version of Eq. (\re{ef})
\be\underline{F_{LM}}(u,v,t)=
\sum_{l_1l_2}
\underline{Y^{l_1l_2}_{LM}}(t)f_{l_1l_2L}(u,v)\la{emod}\ee
since, with the choice made of the body reference frame, the functions
$\underline{Y^{l_1l_2}_{LM}}$ entering (\ref{emod})  are also real.

\section{Matrix Elements of NN Interaction and Kinetic Energy}

The basis functions are antisymmetric with respect to particle
permutations so it is sufficient to calculate the  ME of an interaction
between a pair of particles.
We consider here such an interaction, e.g. $\hat{V}(12)$, that includes the
operators
\ber
\hat{l}^2,\quad \hat{k}^2, 
\quad({ \hat{\vec l}}\cdot({ \hat{\vec s}}_1+{ \hat{\vec s}}_2)),\quad
S_{12}=3({\vec n}\cdot\mbox{$\vec{\sigma}$}_1)({\vec n}
\cdot\mbox{$\vec{\sigma}$}_2)
-(\mbox{$\vec{\sigma}$}_1\cdot\mbox{$\vec{\sigma}$}_2),\nonumber\\
S_{12}^{ll}=3({ \hat{\vec l}}\cdot\mbox{$\vec{\sigma}$}_1)({ \hat{\vec l}}
\cdot\mbox{$\vec{\sigma}$}_2)
-\hat{l}^2(\mbox{$\vec{\sigma}$}_1\cdot\mbox{$\vec{\sigma}$}_2),\quad
S_{12}^{kk}=3({ \hat{\vec k}}\cdot\mbox{$\vec{\sigma}$}_1)({ \hat{\vec k}}
\cdot\mbox{$\vec{\sigma}$}_2)
-\hat{k}^2(\mbox{$\vec{\sigma}$}_1\cdot\mbox{$\vec{\sigma}$}_2)\la{oper}
\eer
times factors $V(r)$ and isospin and spin projection operators.
Here ${\vec n}={\vec r}/r$, ${\vec r}={\vec r}_2-{\vec r}_1$,
${ \hat{\vec k}}=(1/2)({ \hat{\vec k}}_2-{ \hat{\vec k}}_1)=
-\imath\partial/\partial {\vec r}$,
${\hat{\vec l}}={\vec r}\times{ \hat{\vec k}}$,
and $\mbox{$\vec{\sigma}$}_i$ are the Pauli matrices.
Most frequently used versions of the realistic NN
interaction are of this form. We shall use Eq. (\re{gen}) to 
obtain ME of the above operators integrated
over Euler angles. Initial ME can  alternatively  be calculated
using Eqs. (\re{ef}), (\ref{exp})
but the 
first of these procedures is 
faster.

Within e.g. the approximation
of equal proton and neutron mass (denoted with $m$)
ME of kinetic energy are expressed in terms of those of the $\hat{k}^2$
operator.
Indeed, kinetic energy $\hat{T}$ of $A$ nucleons
in the center--of--mass system can be written
as $(2mA)^{-1}\sum_{i<j}({\vec \hat{p}}_i-{\vec \hat{p}}_j)^2$, 
where ${\vec \hat{p}}_i$
are the nucleon momenta. Then ME of $\hat{T}$ 
can obviously be calculated as
\begin{eqnarray*} \langle G|\hat{T}|F\rangle=
(A-1)(4m)^{-1}\langle G|({\vec \hat{p}}_1-{\vec \hat{p}}_2)^2|F\rangle 
=(A-1)(\hbar^2/m)\langle G|\hat{k}^2|F\rangle. \end{eqnarray*}

Final expressions for the Euler angle integrated ME of the
operators (\re{oper}) are given by Eqs. (\re{ls}), (\re{l2}), (\re{k2}),
(\re{n}), (\re{l}), (\re{p}), and (\re{fin}) below.
ME of the $ls$ force and of further operators in  (\re{oper})
are expressed in the usual way \cite{var} in terms of the reduced ME
of spatial components of those operators.   
In the $ls$ force case, the corresponding Euler angle
integrated reduced ME $\left(G_{L'}||V(r)\hat{l}||F_{L}\right)$
of the operator $V(r)\hat{l}_\mu$ are required.
Equation (\re{gen}) gives
\be
\left(G_{L'}||V(r)\hat{l}||F_{L}\right)=8\pi^2(2L'+1)^{-1/2}
\int d\tau_{int}V(r)\sum_{MM',\mu=\pm 1\atop M'=M+\mu}C_{LM1\mu}^{L'M'}
\underline{G_{L'M'}}\,\,
\underline{\hat{l}_\mu F_{LM}}.\la{ls}
\ee
Here the
quantities  $\underline{\hat{l}_\mu F_{LM}}$ are defined as above.
One has  $\underline{\hat{l}_0 F_{LM}}=0$ which is  
taken into account in (\re{ls}).
For $\mu=\pm 1$ these quantities are of the form 
\be
\underline{\hat{l}_{\pm 1} F_{LM}}=
-\frac{u}{\sqrt{2}}\left
(\frac{\partial}{\partial u_x'}\pm \imath\frac{\partial}{\partial u_y'}\right)
F_{LM}\la{der}
\ee
where the derivatives are taken at the space point from (\re{poi}).  
The derivatives can be calculated numerically, and high accuracy
is attained in this way.
To perform this for symmetrized basis functions 
one needs to apply Eq. (\re{symm}) at space
points $({\vec u},{\vec v})$ that are obtained from Eq. (\re{poi})
by the replacement of $u_x'=0$, or $u_y'=0$, with small non--zero values
of these quantities.

Only
real parts are to be collected in the right--hand
side of (\re{der}) when the functions $f_{l_1l_2L}$ are real in Eq. (\re{ef}).
Indeed, the quantities $\underline{\hat{l}_{\pm 1}F_{LM}}$
are real in this case which is seen using the expansion 
of $\hat{l}_{\pm 1}F_{LM}$
over $Y_{LM}^{l_1l_2}$.

When calculating ME of the $\hat{l}^2$ operator it is convenient to 
express them in terms
of the quantities that already appear in ME (\re{ls}). 
(The same is done below also in
Eqs. (\re{k2}) and (\re{l}).)
For this purpose one writes down the integrand
in the form $V(r)\sum_\mu (\hat{l}_\mu G_{LM})^*(\hat{l}_\mu F_{LM})$, expresses 
$\hat{l}_\mu F_{LM}$ and $\hat{l}_\mu G_{LM}$ in 
terms
of the Eq. (\re{co}) type quantities, 
uses subsequently Eqs. (\re{rot1}) and (\re{ov}), and
uses expressions  of Eq. (\ref{co1}) type
and the orthogonality property of  
Clebsch--Gordan coefficients.
This gives
\ber
\int d\tau  G_{L'M'}^*V(r)l^2 F_{LM}=\nonumber\\\delta_{LL'}\delta_{MM'}8\pi^2(2L+1)^{-1}
\int d\tau_{int}\,V(r)\sum_{M'',\,\mu=
\pm 1}\underline{\hat{l}_\mu G_{LM''}}\,\,\underline{\hat{l}_\mu F_{LM''}}. \la{l2}
\eer

The Euler angle integrated ME of the $\hat{k}^2$ operator are obtained as
\ber
\int d\tau G_{L'M'}^*\hat{k}^2 F_{LM}=(1/2)\int d\tau G_{L'M'}^*\Delta_uF_{LM}
\nonumber\\=\frac{\delta_{LL'}\delta_{MM'}}{2}
\int d\tau \left(\frac{\partial G_{LM}^*}{\partial u}
\frac{\partial F_{LM}}{\partial u}+G_{LM}^*\frac{\hat{l}^2}{u^2}F_{LM}
\right)=\frac{\delta_{LL'}\delta_{MM'}4\pi^2}{2L+1}\nonumber\\
\times\int d\tau_{int}
\Bigl\{\sum_{M''}[(\partial  /\partial u)\underline{G_{LM''}}\,]
[(\partial  /\partial u)\underline{F_{LM''}}\,]
+u^{-2}
\sum_{M'',\,\mu=\pm 1}\underline{l_\mu G_{LM''}}\,\,\underline{l_\mu F_{LM''}}
\Bigr\}.\la{k2}
\eer
When ME of the operator $1/2[V(r)\hat{k}^2+\hat{k}^2V(r)]$  are considered
the integrand in the last line of (\ref{k2}) is 
obviously to
be replaced with
\begin{eqnarray*}
V(r)\Bigl\{\sum_{M''}[(\partial  /\partial u)\underline{G_{LM''}}\,]
[(\partial  /\partial u)\underline{F_{LM''}}\,]
+u^{-2}
\sum_{M'',\,\mu=\pm 1}\underline{l_\mu G_{LM''}}\,\,\underline{l_\mu F_{LM''}}
\Bigr\}\\
+2^{-1/2}V'(r)\sum_{M''}\Bigl\{\underline{G_{LM''}}
[(\partial  /\partial u)\underline{F_{LM''}}\,]
+[(\partial  /\partial u)\underline{G_{LM''}}\,]
\underline{F_{LM''}}\Bigr\}.
\end{eqnarray*}
The quantities $S_{12}$, $S_{12}^{ll}$, and $S_{12}^{kk}$ in (\re{oper})
can be written as
\[\sum_{i,j=1}^3X_{ij}^{(2)}\Sigma_{ij}^{(2)}=
(3/2)\sum_{\mu=-2}^2X_\mu^{(2)}\Sigma_{-\mu}^{(2)}(-1)^\mu\]
where $X_{ij}^{(2)}$ and $\Sigma_{ij}^{(2)}$ are the Cartesian
components of a space tensor and spin 
 tensor, respectively, and $X_\mu^{(2)}$ and $\Sigma_\mu^{(2)}$
are the corresponding spherical components. The latter are
linear combinations \cite{var} of the Cartesian ones, and in particular
$X_0^{(2)}=X_{zz}^{(2)}$,  $\Sigma_0^{(2)}=\Sigma_{zz}^{(2)}$. In our case
\[\Sigma_{ij}^{(2)}=
(1/2)(\sigma_{1i}\sigma_{2j}+\sigma_{1j}\sigma_{2i})-(1/3)\delta_{ij}
(\mbox{$\vec{\sigma}$}_1\cdot\mbox{$\vec{\sigma}$}_2).\]
One thus has
\begin{eqnarray*}
 S_{12}=
 \frac{3}{2}\sum_{\mu=-2}^2 \mbox{$\cal{N}$}_\mu^{(2)}
 \Sigma^{(2)}_{-\mu}(-1)^\mu,\\
S_{12}^{ll}=\frac{3}{2}\sum_{\mu=-2}^2 
\mbox{$\cal{L}$}_\mu^{(2)}\Sigma^{(2)}_{-\mu}(-1)^\mu,\qquad
S_{12}^{kk}=\frac{3}{2}\sum_{\mu=-2}^2 
\mbox{$\cal{K}$}_\mu^{(2)}\Sigma^{(2)}_{-\mu}(-1)^\mu.\end{eqnarray*}
The tensors $\mbox{$\cal{N}$}_\mu^{(2)}$,
$\mbox{$\cal{L}$}_\mu^{(2)}$, and $\mbox{$\cal{K}$}_\mu^{(2)}$ are
\ber
\mbox{$\cal{N}$}_\mu^{(2)}=4(\pi/5)^{1/2}Y_{2\mu}({\vec n}),\nonumber\\
\mbox{$\cal{L}$}_\mu^{(2)}=
4(\pi/5)^{1/2}\{\mbox{$\cal{Y}$}_{2\mu}({ \hat{\vec l}})\}_{sym},\qquad
\mbox{$\cal{K}$}_\mu^{(2)}=
4(\pi/5)^{1/2}\mbox{$\cal{Y}$}_{2\mu}({ \hat{\vec k}}),
\la{tens}
\eer
where $\mbox{$\cal{Y}$}_{2\mu}({\vec x})=x^2Y_{2\mu}(\bar{{\vec x}})$
are the solid
spherical harmonics. The $\mu=0$ components of these tensors are
\be
\mbox{$\cal{N}$}_0^{(2)}=3n_z^2-1,\quad
\mbox{$\cal{L}$}_0^{(2)}=3\hat{l}_z^2-\hat{l}^2,\quad
\mbox{$\cal{K}$}_0^{(2)}=3\hat{k}_z^2-\hat{k}^2.\la{zero}
\ee
 
To perform integration over Euler angles  in the reduced
ME of the first of the 
tensors (\re{tens}) one
uses Eq. (\re{gen}). One
has $\underline{\mbox{$\cal{N}$}_\mu^{(2)}}=2\delta_{\mu 0}$ that gives
\be
\left(F_{L'}||V(r)\mbox{$\cal{N}$}^{(2)}||G_{L}\right)=16\pi^2(2L'+1)^{-1/2}
\int d\tau_{int}V(r)\sum_{M}C_{LM20}^{L'M}\underline{F_{L'M}}\,\,
\underline{G_{LM}}.\la{n}
\ee

The Euler angle integrated reduced ME of the two other tensors (\re{tens}) 
will be 
obtained
via calculating the corresponding ME
of the operators (\re{zero}).
Proceeding in the same way as at the derivation of Eq. (\re{l2}) we 
come to the following expressions
\ber
\left(G_{L_2}||V(r)\mbox{$\cal{L}$}^{(2)}||F_{L_1}\right)=
\nonumber\\
8\pi^2\int V(r)\,d\tau_{int}\sum_{\mbox{$\mu_1=\pm 1,M_1,\mu_2=
\pm 1,M_2\atop \mu_1+M_1=\mu_2+M_2$}}f(L_1,L_2,M_1,M_2,\mu_1,\mu_2)\,
\underline{l_{\mu_2}G_{L_2M_2}}\,\,\underline{l_{\mu_1}F_{L_1M_1}},\la{l}\\
\left(G_{L_2}||(1/2)[V(r)\mbox{$\cal{K}$}^{(2)}+
\mbox{$\cal{K}$}^{(2)}V(r)]||F_{L_1}\right)=
\nonumber\\
8\pi^2\int d\tau_{int}\sum_{\mbox{$\mu_1M_1\mu_2M_2\atop \mu_1+M_1=
\mu_2+M_2$}}f(L_1,L_2,M_1,M_2,\mu_1,\mu_2)\left
[V(r)\underline{\nabla_{\mu_2}G_{L_2M_2}}\,\,
\underline{\nabla_{\mu_1}F_{L_1M_1}}
\nonumber\right.\\
+2^{-1/2}V'(r)\left.\left(\delta_{\mu_10}
\underline{\nabla_{\mu_2}G_{L_2M_2}}\,\,\underline{F_{L_1M_1}}+
\delta_{\mu_20}\,
\underline{G_{L_2M_2}}\,\,\underline{\nabla_{\mu_1}F_{L_1M_1}}\right)\right]
\la{p}
\eer
which contain the quantity
\ber
f(L_1,L_2,M_1,M_2,\mu_1,\mu_2)=
\frac{(2L_2+1)^{1/2}}{C_{L_1M20}^{L_2M}}
S(L_1,L_2,M_1,M_2,M,\mu_1,\mu_2),\la{f}\\
S(L_1,L_2,M_1,M_2,M,\mu_1,\mu_2)=\nonumber\\
3\left(\sum_{L'}(2L'+1)^{-1}C^{L'M}_{L_1M10}
C^{L'M}_{L_2M10}
C^{L'M'}_{L_1M_11\mu_1}C^{L'M'}_{L_2M_21\mu_2}\right)
-\frac{\delta_{\mu_1\mu_2}\delta_{M_1M_2}
\delta_{L_1L_2}}{2L_1+1}.\la{sum}
\eer
It is implied  here that $|L_1-L_2|\le 2$. The value of $M$ in (\re{f}) 
is arbitrary 
provided
that $C_{L_1M20}^{L_2M}\ne 0$. In (\re{sum}) $M'=\mu_1+M_1=\mu_2+M_2$.
In (\re{p}) $\mbox{${\vec{\nabla}}$}=\partial/\partial {\vec r}$.
The quantities $\underline{\nabla_{\mu}F_{LM}}$ and 
$\underline{\nabla_{\mu}G_{LM}}$
are real at the same conditions as above.

The summation in (\re{sum}) can be done in a closed form. For this
purpose, let us perform a transformation of the sum. It
follows from the general structure of the
ME \cite{var} that the equality
$S=C_{L_1M20}^{L_2M}S'$ with $S'$ being independent of $M$
holds true. Then multiplying $S$ by $C_{L_1M20}^{L_2M}$, summing over $M$
and using the relation $\sum_M\left(C_{L_1M20}^{L_2M}\right)^2=(2L_2+1)/5$
one concludes that the quantity
(\re{f}) may be represented as
$5(2L_2+1)^{-1/2}S_0$, where $S_0=\sum_M C_{L_1M20}^{L_2M}S$.
And due to the relation $\sum_MC_{L_1M20}^{L_2M}=0$ the last term from
(\re{sum}) is eliminated in the course of the summation. As a result, one
obtains
\ber
f(L_1,L_2,M_1,M_2,\mu_1,\mu_2)
=15(2L_2+1)^{-1/2}\sum_{L'M'}g(L_1,L_2,L')C_{L_1M_11\mu_1}^{L'M'}
C_{L_2M_21\mu_2}^{L'M'},
\la{s}\\
g(L_1,L_2,L')=(2L'+1)^{-1}
\sum_M C_{L_1M20}^{L_2M}C_{L_1M10}^{L'M}C_{L_2M10}^{L'M}.\la{g}
\eer
(The summation over $M'$ in (\re{s}) is a formal one.)
 Now we perform the summation in
 (\re{g})
with the help of Eq. 8.7.(13) in \cite{var}
obtaining
\[ g(L_1,L_2,L')=(-1)^{L_1+L'}(2L_2+1)^{1/2}\sqrt{\frac{2}{15}}
\left\{\begin{array}{ccc}
1 & 1& 2\\L_1 & L_2 & L'\end{array}\right\}.\]
We substitute this into (\re{s}) and perform the summations
with the help of Eq.~8.7.(32) in \cite{var}. 
Finally, we obtain  that in (\re{l}), (\re{p})
\be
f(L_1,L_2,M_1,M_2,\mu_1,\mu_2)=(6/5)^{1/2}(-1)^{\mu_1+M_2}
C^{2,M_1-M_2}_{L_1M_1L_2,-M_2}C^{2,\mu_1-\mu_2}_{1\mu_11,-\mu_2}.\la{fin}
\ee
It is implied here that $\mu_1+M_1=\mu_2+M_2$,  otherwise
$f=0$.
 
The formulae in this section have been verified in three--nucleon computations.

\section{Symmetry relations}

We write down the formulae for (components of) states
possessing definite parity $(-1)^{I}$. The relation 
\be
\underline{F_{L,-M}}=
(-1)^{I+L+M}\underline{F_{LM}}\la{sy}
\ee
holds true.
According to (\re{sy}) $\underline{F_{L,M=0}}$ vanishes when $I+L$ is odd.
To obtain (\re{sy}) one notes that in Eq. (\re{emod})
$\underline{Y^{l_1l_2}_{L,-M}}=(-1)^{l_1+l_2+L+M}\underline{Y^{l_1l_2}_{LM}}$.

Let us assume that the quantities
$\hat{O}_{k\kappa}F_{LM}$ have definite parity $(-1)^{i}$ 
(that is related to the parity of $F_{LM}$ and that of the operator).
Then
\be
\underline{\hat{O}_{k,-\kappa}F_{L,-M}}=(-1)^{i+L+M+k+\kappa}
\underline{\hat{O}_{k\kappa}F_{LM}}.\la{ss}
\ee
Eq. (\ref{ss}) is
obtained expressing
its left--hand side in terms  of the functions
$\underline{(OF)_{L_0M_0}^{kL}}$ from 
Eq. (\re{co1}) and applying Eq. (\re{sy})
to them.

The integrands in Eqs. (\re{ov}), (\re{exp}), (\re{n}), and (\re{l2}) 
may be written, up to the
notation of summation, as
\[\sum_M X(M)\qquad \mathrm{and}\qquad \sum_{M\mu} X(M,\mu).\]
Due to Eqs. (\re{sy}) and (\re{ss}) one has
\[X(-M)=(-1)^{I+I'}X(M),\qquad X(-M,-\mu)=(-1)^{I+I'}X(M,\mu),\]
where $(-1)^{I}$ and $(-1)^{I'}$ are the parities of 
the two states entering an ME. 
Then, when parities are the same, one can apply e.g. the relations 
\[\sum\limits_MX(M)=X(0)+2\sum_{M>0} X(M),
\quad \sum\limits_{M\mu}X(M,\mu)=\sum_\mu X(0,\mu)+2\sum\limits_{M>0,\mu} X(M,\mu)\]
to speed up the summations. When parities are opposite
the sums vanish as it should be.

The integrands in Eqs. (\re{sv2}), (\re{ls}), (\re {l}),  and (\re {p}) 
may be written as
\[\sum\limits_{MM'\mu=\pm 1\atop M'=M+\mu}X(M,M',\mu)\qquad \mathrm{and}\qquad 
\sum\limits_{M_1\mu_1M_2\mu_2\atop M_1+\mu_1=M_2+\mu_2}X(M_1,\mu_1,M_2,\mu_2).\] 
One has, with the same definitions of $(-1)^{I}$ and $(-1)^{I'}$ as above,
\begin{eqnarray*}X(-M,-M',-\mu)=(-1)^{I+I'}X(M,M',\mu),\\
X(-M_1,-\mu_1,-M_2,-\mu_2)=(-1)^{I+I'}X(M_1,\mu_1,M_2,\mu_2).\end{eqnarray*}
When one of the summation variables above is expressed
in terms of the others  this leads to the sums of the form
\[\sum_{m_1m_2} Y(m_1,m_2),\qquad \sum_{m_1m_2m_3} Y(m_1,m_2,m_3)\]
with the similar property $Y(-m_1,\ldots,-m_i)=(-1)^{I+I'}Y(m_1,\ldots,m_i)$.
Then, when parities are the same, 
at a particular choice of summation variables 
one  can apply e.g. the relations of the type
\begin{eqnarray*}\sum_{M\mu} Y(M,\mu)=\sum_\mu Y(0,\mu)+2\sum_{M>0,\mu} Y(M,\mu),\\
\sum\limits_{M_1\mu_1\mu_2}Y(M_1,\mu_1,\mu_2)=\sum_{\mu_1\mu_2} Y(0,\mu_1,\mu_2)+
2\sum\limits_{M_1>0,\mu_1\mu_2}Y(M_1,\mu_1,\mu_2).
\end{eqnarray*}
\begin{acknowledge}
The work was partially supported by the Russian Foundation for Basic Research
(grant 00-15-96590).
\end{acknowledge}
\appendix
\addtocounter{section}{1}
\section*{Appendix}
\setcounter{section}{1}

The known relationship $d\bar{{\vec u}}d\bar{{\vec v}}=d\omega dt$ is
obtained as follows. Let us recall the definition, see \cite{var}, of the Euler
angles $\omega\equiv\{\alpha,\beta,\gamma\}$: the angles $\alpha$ and $\beta$
are the azimuthal angle and polar angle determining the direction of the
vector ${\vec u}$, and the angle $\gamma$ corresponds to a rotation
of an intermediate reference frame around its $z$--axis coinciding
with the ${\vec u}$ direction. (This  
rotation brings the $x$--axis of the intermediate reference frame
into the ${\vec u},{\vec v}$ plane.) Let us perform the integration over
$d\bar{{\vec v}}$ choosing some fixed reference frame with the $z$--axis
directed along ${\vec u}$. Then one can write 
$d\bar{{\vec u}}d\bar{{\vec v}}=d\alpha\sin\beta d\beta dtd\varphi$ where
$\varphi$ is the azimuthal angle of the vector ${\vec v}$ in the chosen
reference frame. But, by the definition of $\gamma$, $\varphi$ differs from
$\gamma$ only by a constant, so that $d\varphi=d\gamma$. This gives the
desired expression taking into account that \cite{var} 
$d\omega=d\alpha\sin\beta d\beta d\gamma$.

The D--function orthogonality relations we use are \cite{var}
\be\int d\omega D^{L_2*}_{M_2M_2'}(\omega)D^{L_1}_{M_1M_1'}(\omega)=
\frac{8\pi^2}{2L_1+1}\delta_{L_1L_2}\delta_{M_1M_2}\delta_{M_1'M_2'}.\la{aor}\ee

To obtain (\re{v2}) we need to expand the states of
the form
\[ \left(\hat{O}_{k}\cdot\hat\Sigma_{k}\right)
|\left(F_{l}\otimes\varphi_{s}\right)_{JM}\rangle\]
over those of the form
\[|\left((OF)_L^{lk}\otimes(\Sigma\varphi)_S^{sk}\right)_{JM}\rangle.\]
The expansion coefficients are the same as if $\hat{O}_{k}$ and
$\hat\Sigma_{k}$ were not operators but states of independent
subsystems with angular momenta $k$. Hence, these coefficients are
the recoupling coefficients for four angular momenta \cite{var}, 
i.e. basically $9j$--symbols. Our initial states include
scalar products so that two of the  four angular momenta are coupled
to zero angular momentum. Therefore, the $9j$--symbols reduce 
to $6j$--symbols
in our case.

\end{document}